\documentclass[superscriptaddress,aps,showkeys,nofootinbib,twocolumn]{revtex4-2}

\usepackage[T1]{fontenc}
\usepackage{newtxtext}
\usepackage{newtxmath}
\usepackage{soul}

\usepackage{amsmath}
\usepackage{mathtools}
\usepackage{amsfonts}
\usepackage{dsfont}

\usepackage{color}
\usepackage{xcolor}
\usepackage[colorlinks,citecolor=blue,linkcolor=blue,urlcolor=blue]{hyperref}

\usepackage{graphicx}
\usepackage[all]{hypcap} 

\usepackage[percent]{overpic}
\usepackage{empheq}
\usepackage[most]{tcolorbox}

\begin{document}

\title{Ergotropy of macroscopic classical systems}
\author{Michele Campisi}
\affiliation{Istituto Nanoscienze-CNR, NEST-Scuola Normale Superiore, 56127 Pisa, Italy}
\email{michele.campisi@nano.cnr.it}

\begin{abstract}
We address the fundamental question of what is the maximal amount of energy (a.k.a.  ergotropy) that can be extracted from 
a macroscopic classical system by means of a cyclical variation of an external parameter. 
We show that if the system is prepared in a statistical state of the form  $\rho = F(H)$, with $H$ the Hamiltonian and $F$ a generic limited function, then the ergotropy is vanishing in the {large $N$} limit, provided  the thermodynamic entropy obeys the scaling $\mathcal S(E) \simeq  N^\alpha s ({E}/{N^\beta})$ with $\alpha,\beta>0$ and  $s$ some strictly increasing function with certain reasonable properties. We stress that such ubiquitous condition also ensures the equivalence of the so-called "surface" and "volume" entropies. We illustrate this by introducing the notion of ``ergotropic rearrangement'' of a phase space density which we then apply to find the exact ergotropy expression of systems with constant (positive or negative) heat capacity being evenly distributed over energy shells in phase space.  
We discuss cases where the above mentioned equivalence does (does not) break and the ergotropy does (does not) vanish accordingly, in the large $N$ limit. The presented results shed new  light on the statistical-mechanical foundations of the second law of thermodynamics.
\end{abstract}

\maketitle

\section{Introduction}

What is the maximal amount of energy (a.k.a. ergotropy \cite{Allahverdyan04EPL67}) one can extract from a macroscopic isolated system via a cyclic time-dependent perturbation?

This is an important question, not only from a practical point of view, but especially, from a foundational point of view. According to the Kelvin formulation of the second law of thermodynamics, such maximal energy should be zero if the system is at equilibrium \cite{Allahverdyan02PHYSA305}. 

It is a very well established fact that if the system is the thermal Gibbs equilibrium, one cannot extract (on average) any energy. This is known to be the case for quantum systems \cite{Pusz78CMP58,Lenard78JSP19,Tasaki00arXivb,Allahverdyan02PHYSA305} as well as for classical systems \cite{Gorecki80LMP4,Daniels81JMP22}, regardless of their size. In fact it is well known that this is generally the case for any stationary state of the  form $\rho=g(H)$ with $H$ the system unperturbed Hamiltonian, and $g$ any decreasing function \cite{Pusz78CMP58,Lenard78JSP19,Tasaki00arXivb,Allahverdyan02PHYSA305,Gorecki80LMP4,Daniels81JMP22}.

Hal Tasaki \cite{Tasaki00arXivb} and soon after Allahverdyan and  Nieuwenhuizen \cite{Allahverdyan02PHYSA305} argued that, in the thermodynamic limit, the ergotropy of a system being in a microcanonical equilibrium should vanish, too 
(note that in the microcanonical case it is  $\rho=F(H)$, but with $F$ being a delta function, i.e., not a decreasing function). This issue was later investigated quantitatively for the prototypical case of a quantum system composed of $N$ independent spin $1/2$ particles \cite{Allahverdyan11EPL95}. The study evidenced a sub-extensive, rather than vanishing, behaviour of the ergotropy of microcanonical states.

Here we set out to shed light on this fundamental issue, by investigating the ergotropy of macroscopic classical systems. 
Our main result is that any equilibrium of the form $\rho=F(H)$ with a generic (non-necessarily decreasing, or smooth) $F$ is asymptotically passive 
({namely has vanishing ergotropy in the large $N$ limit}), provided a proper notion of thermodynamic limit applies, whose validity is most general and ubiquitous, see Eq. (\ref{eq:exponential-volume}) below.
This is illustrated by providing the analytical expression of ergotropy of systems with either positive or negative constant heat capacity.

\subsection{Summary of results}

Our main tool of investigation is what we call the shell-ergotropy, namely the ergotropy of a flat distribution whose support is an energy shell in phase space.
In order to study the shell-ergotropy, we build on the recent work of  \cite{Campisi26EPL154} presenting the fundamentals of classical ergotropy and an analytical formula for its calculation, see Eq. (\ref{eq:erg}) below. However said formula is valid in the case of phase-space densities $\rho$ that are continuous and without flat plateaus, which is not the case of distributions defined on a shell.

Our first result consists in finding a general formula, see Eq. (\ref{eq:ergotropic-rearrangement}) below, that generalises Eq. (\ref{eq:erg}), without restrictions on its applicability. We achieve such formula  by establishing a connection with the mathematical literature: The problem of classical ergotropy, is, mathematically speaking, analogous to the problem of ``symmetric decreasing rearrangement'' (see, e.g., Lieb and Loss \cite{Lieb01Book}). Its generalisation to possibly non-symmetric rearrangements, whereby a distribution is rearranged in a way that it decreases as some other function (specifically the energy) increases, leads to the notion of ``ergotropic rearrangement", which, in turn, leads to a most general expression of classical ergotropy.

Our second result is the explicit expression of the shell-ergotropy of a classical ideal gas, Eq. (\ref{eq:erg+}), and the finding that, in accordance with Refs. \cite{Tasaki00arXivb,Allahverdyan02PHYSA305} it indeed tends to vanish as the number of particles increases. In fact we establish that it does so exponentially, when all other relevant quantities are being kept fixed. We identify the mechanism at the basis of the vanishing of shell-ergotropy as the phenomenon of concentration of measure \cite{Levy51Book}. 

Since concentration of measure is one of the mechanisms at the basis of the equivalence of canonical and microcanonical ensembles \cite{Huang87Book} one might be tempted to guess that such equivalence is necessary for the shell-ergotropy to asymptotically vanish. 
Our third result is that this is not the case. We report the explicit expression of shell-ergotropy, Eq. (\ref{eq:erg-}) for the prototypical case where the equivalence of microcanonical and canonical ensembles break, namely systems with constant negative microcanonical heat capacity\footnote{Heat capacity cannot get negative values in the canonical ensemble \cite{Campisi25Book}. For sake of simplicity, here we shall often use the expression ``negative heat capacity'' instead of  ``negative microcanonical heat capacity''.} \cite{Thirring70ZPA235,Campisi12EPL99}  and find that again the shell-ergotropy exponentially vanishes in the {large $N$} limit.

Our fourth result, see Sec. \ref{subsec:artificial}, consists in providing a (rather artificial) macroscopic model defying the concentration of measure phenomenon. It is obtained by composing  a system with large positive heat capacity with a system with slightly larger (in modulo) system with negative heat capacity. Such a system has a small finite negative heat capacity, despite being composed of an arbirary large number of degrees of freedom, and its shell-ergotropy remains finite accordingly.

Finally we establish that the ergotropy of a generic stationary state of the form $F(H)$ is asymptotically vanishing provided the system admits a proper thermodynamic limit, Eq. (\ref{eq:exponential-volume}). Such  thermodynamic limit holds ubiquitously, not only for standard extensive matter, but also for possibly non-extensive systems (e.g., featuring long-range interactins and/or negative heat capacities). We remark that the existence of such a limit also ensures the equivalence of ``surface'' and ``volume'' entropies \cite{Dunkel14NATPHYS10,Hilbert14PRE90,Campisi15PRE91}.

\section{Quick review of classical ergotropy}

Given a classical system with Hamiltonian $H(\mathbf{z})$ being prepared in a statistical state $\rho(\mathbf{z})$, its ergotropy $\mathcal E$, is defined as the maximal amount of energy that, on average, can be extracted from the system by means of a cyclical time-dependent driving $V(\mathbf{z},t)$. Here $\mathbf{z}=(\mathbf{q},\mathbf{p})$ is a point in the system phase space. In formal terms  
\begin{align}
\label{eq:def-erg}
\mathcal E[\rho] = \sup_{\varphi} \int d\mathbf z H(\mathbf{z})[\rho(\mathbf{z}) - \rho(\varphi^{-1}(\mathbf{z})) ]
\end{align}
where the ``$\sup$'' is taken over all possible Hamiltonian flows $\varphi:\mathbb R^{2f}\to \mathbb R^{2f}$ and $f$ denotes the number of degrees of freedom.\footnote{We point out that the present definition (\ref{eq:def-erg}) differs from the one in Ref. \cite{Campisi26EPL154} in that the symbol ``$\max$'' is here correctly replaced by the appropriate symbol ``$\sup$''. This is because said ``$\sup$'' value can be approached to any wanted to degree of accuracy by a Hamiltonian flow, but it is not guaranteed that it be saturated.}

Reference \cite{Campisi26EPL154} provides an explicit formula for $\mathcal E$ valid in the case where $\rho$ is continuous and does not present flat plateaus. It is based on the observation that any measure preserving map $\mathcal M$ can be approximated to any wanted accuracy by a (possibly non smooth) Hamiltonian flow $\varphi$ \cite{Katok73MathUSSRIzv7,Campisi26EPL154}. That allows to write the ergotropy as a ``$\max$'' over all measure preserving maps 
\begin{align}
\label{eq:def-erg2}
\mathcal E[\rho] = \max_{\mathcal M} \int d\mathbf z H(\mathbf{z})[\rho(\mathbf{z}) - \rho(\mathcal M^{-1}(\mathbf{z})) ]\, .
\end{align}
We shall use the symbol $\breve \rho$ for the state of minimal energy expectation among all possible evolved states $\rho(\mathcal M^{-1}(\mathbf{z}))$, which is known as the passive companion of $\rho$ \cite{Gorecki80LMP4,Daniels81JMP22}.

The explicit expression of ergotropy then follows from Gardner's principles \cite{Gardner63PF6}, according to which i) the passive state $\breve \rho(\mathbf z)$  is of the form $g(H(\mathbf{z}))$ with $g$ a non-increasing function, ii)  the measure of the set $\rho > r$ has to be the same as the measure of the set $\breve \rho > r$, for all $r >0$. Introducing the function
\begin{align}
\Sigma(r)&=  \int d\mathbf z \, \theta[\rho(\mathbf{z})-r]
\label{eq:Sigma}
\end{align}
where $\theta$ is Heaviside step function, the two Gardner principles can be succintly expressed as:
\begin{align}
\Sigma(r) = \int d\mathbf z \, \theta[g(H(\mathbf{z}))-r]\, .
\end{align} 
Assuming $g$ is strictly decreasing, a fact that needs to be checked for consistency later, the above is equivalent to 
\begin{align}
\Sigma(r) &= \int d\mathbf z \, \theta[g^{-1}(r)- H(\mathbf{z})] = \Omega(g^{-1}(r)) \,
\label{eq=Sigma=Omega°g}
\end{align} 
where we introduced the so called ``phase volume''
\begin{align}
\Omega(E) = \int d\mathbf z \, \theta[E-H(\mathbf{z})]
\label{eq:Omega_0}
\end{align}
Equation (\ref{eq=Sigma=Omega°g}) appeared already decades ago in the astrophysics literature \cite{Wiechen88MNRAS232} and more recently in the plasma physics literature \cite{Helander17JPP83}, for the special case of collisionless fluids obeying Vlasov dynamics.

The derivation of Ref. \cite{Campisi26EPL154} proceeds from Eq.  (\ref{eq=Sigma=Omega°g}), by further assuming $\Sigma$ is invertible. Note that by definition $\Sigma$ is non-increasing, so the assumption is equivalent to the assumption that $\Sigma$ is continuous and strictly decreasing, i.e., it has no jumps nor flat regions. The latter can be ensured by assuming $\rho$ does not have jump discontinuities nor flat regions of finite measure. Note in fact that, if $\rho$ has some jump discontinuity, that will result in a flat region of $\Sigma$. Similarly, the presence of a flat region in $\rho$, will result in a jump discontinuity of $\Sigma$.
This will be illustrated below, in Sec. \ref{sec:fate}.

Under the above assumption, rewriting Eq. (\ref{eq=Sigma=Omega°g}) as $\Sigma = \Omega \circ g^{-1}$, and applying $\Sigma^{-1}$ to the left and $g$ to the right, one obtains $g = \Sigma^{-1} \circ \Omega$. Since for any physical Hamiltonian $H$,  $\Omega$ is strictly increasing, and we assumed $\Sigma$ is strictly decresing, $g$ is indeed strictly decreasing, as assumed above. So the derivation is self-consistent, we can write the passive state $\breve \rho$ as 
\begin{align}
\breve \rho(\mathbf z)=\Sigma^{-1}(\Omega(H(\mathbf{z})))
\label{eq:breve-rho}
\end{align}
and obtain
\begin{align}
\mathcal E[\rho] =  \int d\mathbf z H(\mathbf{z})[\rho(\mathbf{z}) - \Sigma^{-1}(\Omega(H(\mathbf{z})))]
\label{eq:erg}
\end{align}

\section{Ergotropic rearrangement of a phase space density}

In order to overcome the limitations of continuity and absence of extended flat regions of $\rho$, we note that the state of minimal energy $\breve \rho$ that one needs to find in order to calculate the ergotropy $\mathcal E$, is, mathematically speaking, a \emph{rearrangement} of $\rho$. 

Classic texts such as that of Hardy, Littlewood and Pólya \cite{Hardy34Book} or the modern text of Lieb and Loss \cite{Lieb01Book}, discuss a special type of rearrangement called ``symmetric decreasing rearrangement'' whereby a  measurable\footnote{Here and in the following by ``measure'' we mean the ``Lebesgue measure''.} non-negative function $\rho:\mathbb  R^f \to \mathbb R_+$  is rearranged in such a way that i) the new function $\rho^*$ does not increase as the distance from the origin increases, that is $\rho^*$ is of the form $\rho^*(\mathbf{z})=g(|\mathbf z|)$ for some non-increasing $g$, ii) the measure of the set where $\rho > r$ is the same as the measure of the set where $\rho^* > r$, for all $r >0$ \cite{Lieb01Book}. 

Clearly, these are the same as Gardner prescriptions, with the only difference that they are formulated for the specific  spherically symmetric function $|\mathbf z|$ instead of a generic (possibly non-spherically symmetric) function $H(\mathbf{z})$.

Let us then quickly review the theory of ``symmetric decreasing rearrangement'', and then formulate its generalisation, leading to the notion of ``ergotropic rearrangement''.

One begins by defining the symmetric rearrangement $A^*$ of a set $A$:
\begin{align}
A^*= \{ \mathbf z : |\mathbf z| \leq R(A)\},  
\end{align}
where $R(A)$ is the radius of the ball that has same measure $\mu(A)$ of the set $A$. Namely, such that:
\begin{align} 
\mu(A)=\mu(A^*)
\end{align}

 Denoting the indicator function of the set $A$ as $\chi_A$, one defines its symmetric rearrangement as:
\begin{align} 
\chi^*_A=\chi_{A^*}^{\phantom{*}}
\end{align}
Finally, recalling the so called layer-cake representation \cite{Lieb01Book}
\begin{align}
\rho(\mathbf z) =  \int_0^\infty dr \chi_{\{\mathbf x: \rho(\mathbf x)>r \}}(\mathbf z)
\end{align}
the symmetric decreasing rearrangement $\rho^*$ of a probability density function $\rho$ is obtained by replacing the indicator functions $\chi$ with their symmetric decreasing rearrangements $\chi^*$ in the above equation, to obtain
\begin{align}
\rho^*(\mathbf z) =  \int_0^\infty dr \chi^*_{\{\mathbf x: \rho(\mathbf x)>r \}}(\mathbf z)
\end{align}
This rearranges the distribution in a measure preserving manner, in such a way that it decreases as the \emph{distance} from the origin increases  \cite{Lieb01Book}. 

The only difference between the symmetric rearrangement $\rho^*$ and the minimal energy state $\breve \rho$ is that the latter is arranged in such a way that it decreases as the \emph{energy}, i.e. $H(\mathbf{z})$, increases, rather than the distance from the origin $|\mathbf z|$. 
In analogy with the above, we then define  the \emph{ergotropic rearrangement} $\breve A$ of a set $A$, as:
\begin{align}
\breve A= \{ \mathbf z : H(\mathbf{z}) \leq  \Omega^{-1}(\mu(A))\} .
\label{eq:breveA}
\end{align}
Note that the the condition $ H(\mathbf{z}) \leq  \Omega^{-1}(\mu(A))$ ensures that 
$
\mu(A)=\mu(\breve A)
$
in fact:
\begin{align}
\mu(\breve A) &= \int d\mathbf{z}\, \theta[\Omega^{-1}(\mu(A))-H(\mathbf{z})] \nonumber \\
&= \Omega(\Omega^{-1}(\mu(A)))=\mu(A)
\end{align}

We then define the ergotropic rearrangement of the indicator functions, as 
\begin{align} 
\breve \chi_A=\chi_{\breve A}
\end{align}
Finally, the ergotropic rearrangement of a  probability density $\rho$ is defined as
\begin{align}
\breve \rho(\mathbf z) =  \int_0^\infty dr \breve \chi_{\{\mathbf x: \rho(\mathbf x)>r \}}(\mathbf{z}).
\label{eq:ergotropic-rearrangement}
\end{align}

The expression in Eq. (\ref{eq:ergotropic-rearrangement}) generalises that of Eq. (\ref{eq:breve-rho}) to the case of a generic phase space density $\rho$. To see that, let us rewrite it in terms of the basic functions $\Sigma$ and $\Omega$, Eqs. (\ref{eq:Sigma},\ref{eq:Omega_0}). First note that the measure of the set $A(r)=\{\mathbf x: \rho(\mathbf x) \geq r \}$ is nothing but $\Sigma(r)$:
\begin{align}
\mu(A(r))= \int d\mathbf z \,\theta( \rho(\mathbf z)-r) =\Sigma(r) 
\end{align}
therefore, its ergotropic rearrangement $\breve A(r)$ reads
\begin{align}
\breve A(r) = {\{\mathbf x: H(\mathbf x)< \Omega^{-1}(\Sigma(r)) \}}
\end{align}
Accordingly, for the indicator function in Eq. (\ref{eq:ergotropic-rearrangement}), we have
\begin{align}
 \breve \chi_{\{\mathbf x: \rho(\mathbf x)>r \}}(\mathbf z) &= \theta[\Omega^{-1}(\Sigma(r))-H(\mathbf{z}) ]\nonumber \\
 &=\theta[\Sigma(r)-\Omega(H(\mathbf{z})) ]
\end{align}
where we used the fact that $\Omega$ is strictly increasing. Plugging this in Eq.  (\ref{eq:ergotropic-rearrangement}) we finally obtain
\begin{align}
\breve \rho(\mathbf z)=\int_0^\infty dr\, \theta[\Sigma(r)-\Omega(H(\mathbf{z})) ]
\label{eq:H-rearrangement}
\end{align}
This is the general expression we were looking for.

To see that Eq. (\ref{eq:breve-rho}) is a special case of Eq. (\ref{eq:H-rearrangement}) note that given a positive, strictly decreasing continuous function $h:\mathbb R_+ \to \mathbb R_+$, its inverse can be written as:
\begin{align}
h^{-1}(y)= \int_0^\infty dx\, \theta [h(x)-y] 
\end{align}
Thus, for a strictly decreasing continuous $\Sigma$, Eq. (\ref{eq:H-rearrangement}) coincides with Eq. (\ref{eq:breve-rho}). 

For a spherically symmetric Hamiltonian, e.g., the quadratic Hamiltonian $H(\mathbf{z})= |\mathbf z|^2$, $\breve \rho$ coincides with $\rho^*$, so we see that ergotropic rearrangement is a generalisation of symmetric decreasing rearrangement.

\section{Shell-ergotropy}
\label{sec:shell}
\begin{figure}[t]
\includegraphics[width=\linewidth]{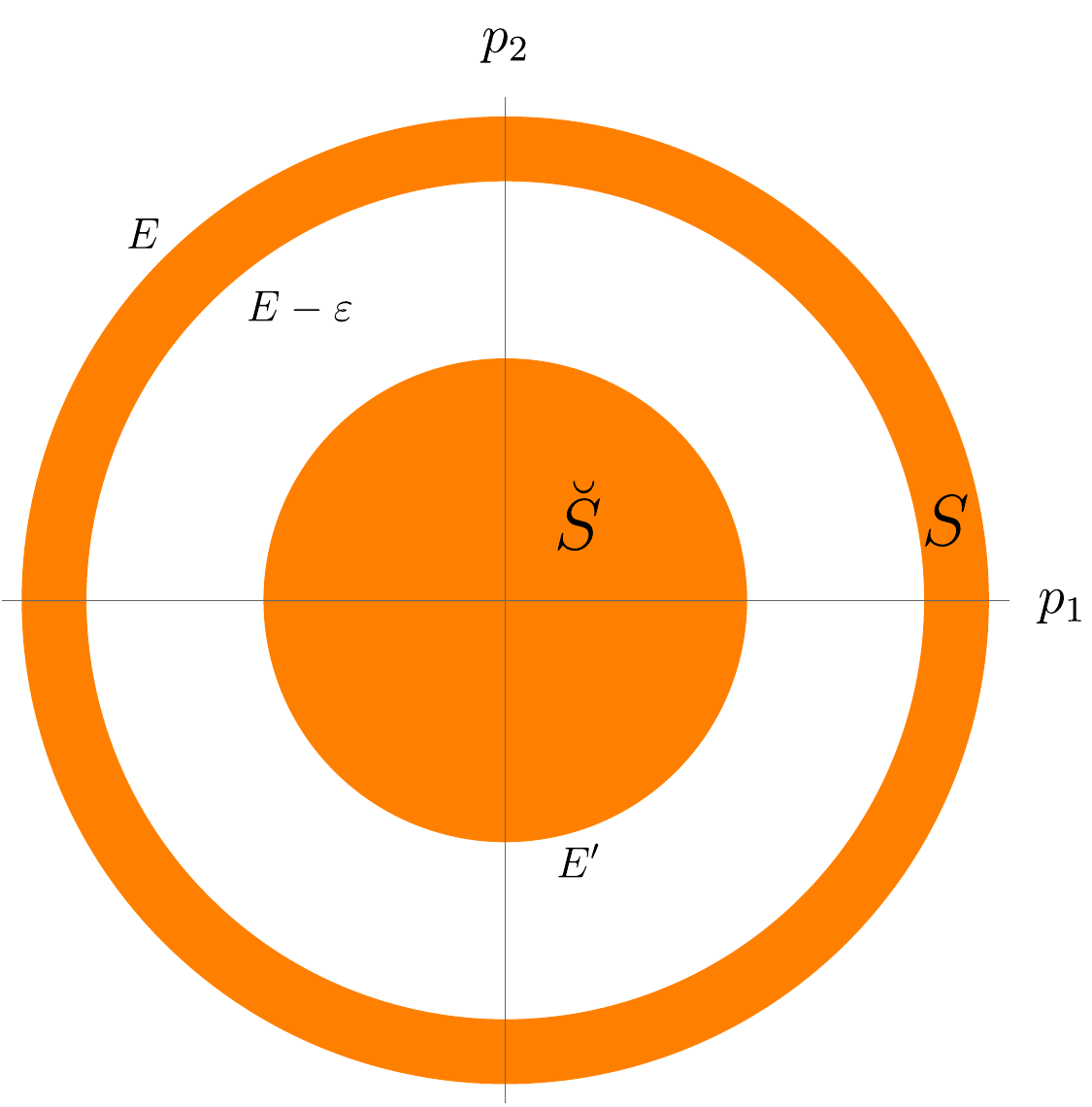}
\caption{The energy shell $S$, and its ergotropic rearrangement $\breve S$. The sets are displayed in the momenta space $(p_1,p_2)$, for the case of only two degrees of freedom. The set $S$ is bounded by the hypersurfaces of energy $E$ and $E-\epsilon$. The set $\breve S$ is bounded by the hypersurface of energy $E'$, Eq. (\ref{eq:E'}). $\rho$ and $\breve \rho$ are evenly distributed on $S$ and $\breve S$, respectively. }
\label{fig:rearrangement} 
\end{figure}

We focus on the case when the system is evenly distributed over an energy shell of finite thickness $\epsilon<E$
\begin{align}
{S}= \{\mathbf z: E-\epsilon \leq H(\mathbf{z}) \leq E \}
\label{eq:S}
\end{align}
that is we study distributions of the type:
\begin{align}
\rho(\mathbf z)= \frac{1}{\mu({S})} \chi_{{S}}(\mathbf z)
\label{eq:rho-in}
\end{align}
which we shall refer to as shell-distributions. 
See Fig. \ref{fig:rearrangement}. Clearly $\rho$ has a jump discontinuity on the border of $S$ and is flat both on the set ${S}$ and out of it.

In the following we shall refer to the ergotropy of shell-distribution, as  the shell-ergotropy.

Using Eq. (\ref{eq:Sigma}) we see that the function $\Sigma(r)$ is constant and equal to $\mu({S})$ for $r<1/\mu({S})$, and is null otherwise:
\begin{align}
\Sigma(r) = \mu({S}) \, \theta[ {1}/{\mu({S})}-r]
\end{align}
Evidently $\Sigma$ has a jump discontinuity and presents extended flat regions, so it is neither continuous nor strictly decreasing. 

Plugging it in Eq. (\ref{eq:H-rearrangement}) and recalling that by definition $\Omega\geq 0$ we swiftly obtain
\begin{align}
\breve \rho(\mathbf z)&= \frac{\theta [\mu({S}) -\Omega(H(\mathbf{z}))]}{\mu({S})}
\label{eq:breverrho1}
\end{align}
Using that $\Omega^{-1}$ is a strictly increasing function, we obtain
\begin{align}
\breve \rho(\mathbf z)&= \frac{\theta [E' -H(\mathbf{z})]}{ \Omega(E)-\Omega(E-\epsilon)}
\end{align}
where we used the explicit expression of $\mu({S})$
\begin{align}
\mu({S}) =  \Omega(E)-\Omega(E-\epsilon)
\label{eq:mu-S}
\end{align} 
and we have introduced the quantity
\begin{align}
\label{eq:E'-def}
E'  \doteq \Omega^{-1}(\mu({S})) =  \Omega^{-1}\left( \Omega(E)-\Omega(E-\epsilon) \right)
\end{align}
representing the bounding energy of the rearranged set $\breve S$, see Fig. \ref{fig:rearrangement}.

The system energy expectation is given by
\begin{align}
\label{eq:initialE}
\bar E&= \int d\mathbf z H(\mathbf{z}) \rho(\mathbf z)
= \frac{1}{\mu(S)} \int_{E-\epsilon}^E de \omega(e) e
\end{align}
where  $\omega(x)=d\Omega(x)/dx$ is the density of states.
The lowest energy the system can reach reads:
\begin{align}
\label{eq:finalE}
\breve E&= \int d\mathbf z H(\mathbf{z}) \breve \rho(\mathbf z) 
= \frac{1}{\mu(S)} \int_{E_0}^{E'} de \omega(e) e 
\end{align}
where $E_0$ is the minimum of $H(\mathbf z)$ over the whole phase space.
Finally, the shell-ergotropy is 
\begin{align}
\mathcal E = \bar E- \breve E = \frac{1}{\mu(S)} \left( \int_{E-\epsilon}^E-  \int_{E_0}^{E'} \right) de\, \omega(e) e
\end{align}

\section{Systems with constant heat capacity}
\label{sec:fate}

\subsection{Ideal gas}
Let us consider a classical ideal gas of $N$ particles in $d$ dimensions. The Hamiltonian reads:
\begin{align}
H(\mathbf q, \mathbf p) = \frac{|\mathbf p|^2}{2m} + U_\text{box}(\mathbf q;V)
\end{align}
where $ U_\text{box}(\mathbf q;V)$ confines the particles to a spatial region of volume $V$ and $\mathbf q= (q_1 \dots q_{dN}), \mathbf p= (p_1 \dots p_{dN})$. The phase volume $\Omega$ is given by \cite{Campisi25Book}
\begin{align}
\Omega(E) = A_\gamma E^{\gamma}, \quad A_\gamma =  \frac{\pi^{\gamma}}{\Gamma(1+\gamma)}(2m)^{\gamma} V^{2\gamma/d}
\label{eq:idealGasVolume}
\end{align}
where $\gamma = dN/2$ is in fact the system (constant) heat capacity. Its inverse is:
\begin{align}
\Omega^{-1}(x) = \left(\frac{x}{A_\gamma}\right)^{1/\gamma} 
\label{eq:idealGasVolumeInverse}
\end{align}
The measure of the shell $S$, Eq. (\ref{eq:mu-S}), is
\begin{align}
\mu({S}) = A_\gamma[E^\gamma-(E-\epsilon)^\gamma]
\end{align}
Using Eqs. (\ref{eq:E'-def},\ref{eq:idealGasVolumeInverse}) we have
\begin{align}
E' = [E^\gamma-(E-\epsilon)^\gamma]^{1/\gamma}
\label{eq:E'}
\end{align}
The system energy expectation, Eq. (\ref{eq:initialE}), is easily calculated as:
\begin{align}
\bar E&= E\frac{\gamma}{\gamma+1}\frac{1-(1-\epsilon/E)^{\gamma+1}}{1-(1-\epsilon/E)^\gamma} 
\end{align}
Using that $E_0=0$, in this case, the lowest energy expectation the system can reach, Eq. (\ref{eq:finalE}), is given by
\begin{align}
\breve E&  =E \frac{\gamma}{\gamma+1} [1-(1-\epsilon/E)^\gamma]^{1/\gamma} 
\end{align}
It is interesting to note that both $\bar E$ and $\breve E $ do not depend on the mass $m$, nor on the volume $V$ of the box.
Taking their difference $\mathcal E = \bar E-\breve E$ we finally get
\begin{align}
\label{eq:erg+}
\mathcal E &= E f_\gamma\left(\frac{\epsilon}{E}\right) \\
f_\gamma(x) &\doteq
\frac{\gamma}{\gamma+1}\left[\frac{1-(1-x)^{\gamma+1}}{1-(1-x)^\gamma} - [1-(1-x)^\gamma]^{1/\gamma} \right]
\label{eq:f-gamma}
\end{align}

\begin{figure}[t]
\includegraphics[width=\linewidth]{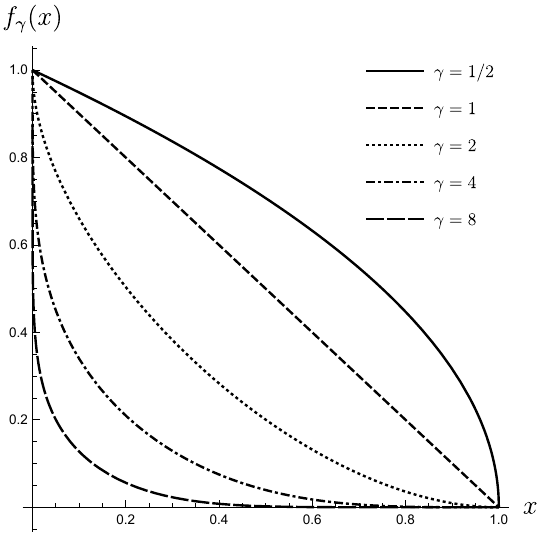}
\caption{Plots of $f_\gamma(x)$, Eq. (\ref{eq:f-gamma}), as function of $x$ for various values of $\gamma$.}
\label{fig:Erg+} 
\end{figure}
\begin{figure}[t]
\includegraphics[width=\linewidth]{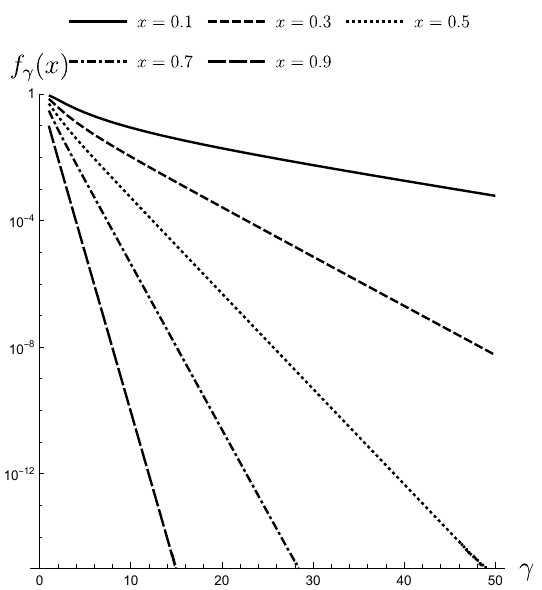}
\caption{Log-plots of $f_\gamma(x)$, Eq. (\ref{eq:f-gamma}), as function of $\gamma$  for various values of $x$.}
\label{fig:Erg+log} 
\end{figure}

Figure \ref{fig:Erg+} shows plots of $f_\gamma(x)$ for different values of $\gamma$. Note how
regardless of the value of $\gamma$, $f_\gamma(0)=1$, implying that the ergotropy is $E$ in the limit of vanishing $\epsilon$. The smaller $\epsilon$, as compared to $E$, the smaller the measure of the set $S$, the more its ergotropic rearrangement $\breve S$ gets concentrated around the origin of momenta space, i.e., the point of zero energy.
Furthermore, regardless of the value of $\gamma$, $f_\gamma(1)=0$, implying that the ergotropy is null in the limit of $\epsilon \to E$. This is because, in this limit $S$ approaches its own ergotropic rearrangement $\breve S$.
Note the special case $\gamma=1$, for which it is $f_1(x)=1-x$, that is $\mathcal E = E-\epsilon$. This linear beahviour is a consequence of the phase volume $\Omega$ being linear in its argument in this case.

Most remarkably, at any fixed value $0<x<1$, $f_\gamma$ tends to vanish as $\gamma$ increases. Figure \ref{fig:Erg+log} shows log-plots of $f_\gamma(x)$ as a function of $\gamma$ for various fixed values of $x$. Note the linear behaviour for large $\gamma$, which evidences an exponential dependence in $\gamma$. This is best understood based on the asymptotic expansion of $f_\gamma(x)$ for large $\gamma $ and $0<x<1$, reading (see appendix \ref{app:asymptotics}):
\begin{align}
\label{eq:f-asymptotic}
f_\gamma(x) \simeq x(1-x)^\gamma \quad (\gamma \gg 1)
\end{align}
Thus for large $\gamma$ we have:
\begin{align}
\mathcal E \simeq \epsilon (1-\epsilon/E)^\gamma\quad (\gamma \gg 1)
\label{eq:f-asymptotic-energy}
\end{align}

The fact that the shell-ergotropy of the ideal gas would vanish in the {large $N$} limit  evidences that despite the state $\rho(\mathbf z)$, Eq. (\ref{eq:rho-in}), is not a passive state for any finite $N$, the energy you can extract therefrom tends to vanish as $N$ increases. This is a consequence of the phenomenon of concentration of measure \cite{Levy51Book}. In the limit of large space dimension, the measure of a ball concentrates on its outer surface. Accordingly, the measure of the shell ${S}$ is all concentrated on its outer surface $H=E$, therefore there is no way to squeeze it in an inner ball of same measure, for that measure, too, will be all concentrated on the ball outer surface.

It is important to stress that if the large $\gamma$ limit is taken while letting $E$ grow linearly with $\gamma$, i.e., $E=\gamma e$, from Eq. (\ref{eq:f-asymptotic-energy}) we get 
\begin{align}
\mathcal E \simeq \epsilon e^{-\epsilon/e}, \quad (E/\gamma= e)
\label{eq:erg-ideal-gas-rescaled}
\end{align}
namely the ergotropy would be of order $1$, i.e., vanishing as compared to the energy $E$ which is of order $\gamma$.

\subsection{Systems with constant negative heat capacity}
\begin{figure}[t]
\includegraphics[width=\linewidth]{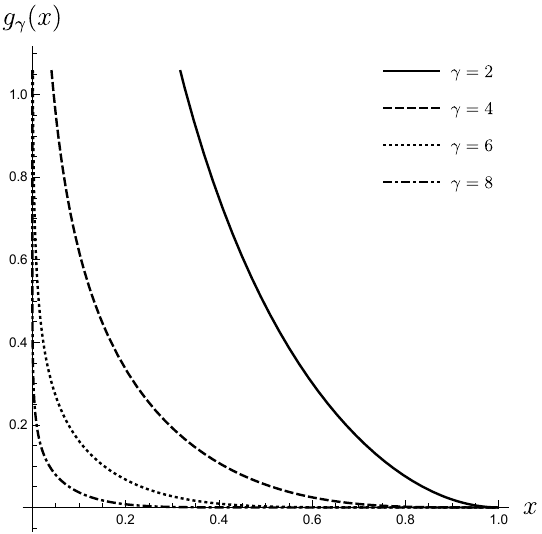}
\caption{Plots of $g_\gamma(x)$, Eq. (\ref{eq:g-gamma}), as function of $x$ for various values of $\gamma$.}
\label{fig:Erg-} 
\end{figure}
\begin{figure}[t]
\includegraphics[width=\linewidth]{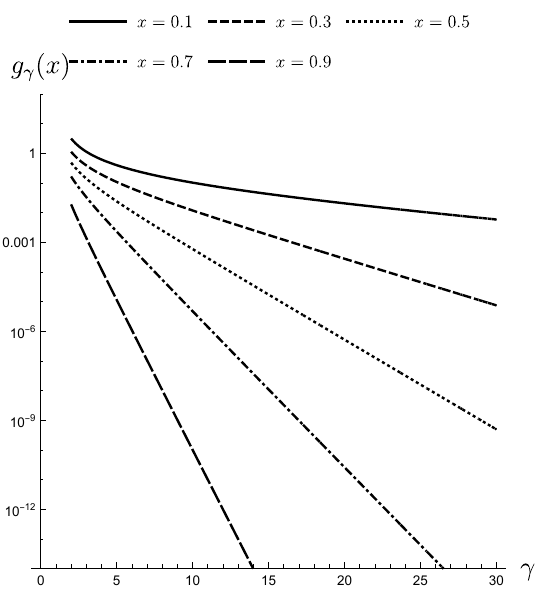}
\caption{Log-plots of $g_\gamma(x)$, Eq. (\ref{eq:g-gamma}), as function of $\gamma$  for various values of $x$.}
\label{fig:Erg-log} 
\end{figure}

Ideal gases have constant positive heat capacity. It is instructive to see how ergotropy behaves for systems with constant \emph{negative} (microcanonical) heat capacity $C=-\gamma<0$, i.e. systems having phase volume of the form \cite{Campisi12EPL99}:
\begin{align}
\Omega(E) =B_\gamma (-E)^{-\gamma} \quad E<0, \gamma>0
\label{eq:Omega-negC}
\end{align}
A  prototypical example is a single particle in 3D subject to the gravitational field, for which $C=-3/2$ \cite{Thirring70ZPA235}. 
In general a single particle in 3D subject to a potential of the form $(x_1^2+x_2^2+x_3^2)^{3/(2C-3)}$ has a phase volume proportional to  $(-E)^{C}$ for $C<0$ \cite{Campisi12EPL99}. It can also be proved that the Hamiltonian:
\begin{align}
H(\mathbf x, \mathbf p) = \sum_{i=1}^f \frac{p_i^2}{2m} - \frac{G}{\sqrt{\sum_{i=1}^f{x_i^2}}}
\label{eq:HnegClarge}
\end{align}
results in a phase volume of the type in Eq.  (\ref{eq:Omega-negC}) where $\gamma = f/2$ with $f$ the number of degrees of freedom (see appendix \ref{app:phase-volume-negC}).

Following  calculations analogous to those illustrated above, the shell-ergotropy is now found to be
 (see appendix \ref{app:derivation-erg-})
\begin{align}
\label{eq:erg-}
\mathcal E &= (-E) g_\gamma\left(\frac{\epsilon}{-E}\right) \quad (\gamma>1)\\
g_\gamma(x) &\doteq 
\frac{\gamma}{1-\gamma} \left[
\frac{(1-x)^{1-\gamma}-1}{(1-x)^{-\gamma}-1} - [(1-x)^{-\gamma}-1]^{-1/\gamma}  \right]
\label{eq:g-gamma}
\end{align}
where now the bounding energies defining the energy shell $S$ are denoted by $E$ and $E+\epsilon$ (with $E<0, \epsilon >0, E+\epsilon<0$).

Figure \ref{fig:Erg-} shows plots of $g_\gamma(x)$ for various values of $\gamma$. A quite interesting feature is that regardless of $\gamma$,  $g_\gamma(1)=0$. This is because this limit corresponds $\epsilon = -E$, namely the upper energy of the set $S$ is $0$. This corresponds to a diverging phase volume, $\Omega \to \infty $, see Eq. (\ref{eq:Omega-negC}), therefore the post-rearrangement bound energy $E'$ would tend to be null too. 
Note also how, regardless of $\gamma$,  $g_\gamma(x)$ diverges for vanishing $x$. This is because for a vanishing $\epsilon$, the measure of $S$ would vanish, and the rearrangement $S$ becomes a ball concentrated around the origin, i.e., at the center of the attractive potential where the energy tends to $-\infty$. 

Like in the ideal gas case, $g_\gamma(x)$ tends to vanish with increasing $\gamma$, at fixed $0<x<1$. This is further illustrated in Fig. \ref{fig:Erg-log} showing log-plots of $g_\gamma(x)$ as function of $\gamma$ for various values of $x$. The large $\gamma$ linear behaviour evidences an exponential dependence of $\gamma$. The asymptotic expansion in the range $0<x<1$ (see appendix \ref{app:asymptotics}) in fact reveals that
\begin{align}
g_\gamma(x) \simeq x(1-x)^\gamma \quad (\gamma \gg 1)
\label{eq:g-asymptotic}
\end{align}
just like in the ideal gas case. Thus for large $\gamma$ it is
\begin{align}
\mathcal E \simeq \epsilon (1+\epsilon/E)^\gamma\quad (\gamma \gg 1)
\end{align}

With this example we have learned that concentation of measure is at play also in the case of systems with constant negative heat capacity, in the limit where such heat capacity is macroscopically large. 

\subsection{An artificial model that defies the concentration of measure}
\label{subsec:artificial}
Is it possible to devise a classical Hamiltonian system such that its shell-ergotropy remains finite as the number of degrees of freedom diverge? 
In other terms, does it exist a classical model defying the concentration of measure phenomenon?
 
One such model can be constructed in the following way. 
First note that when a system with constant negative microcanonical heat capacity $C_1<0$ is combined with a system with constant positive microcanonical heat capacity $C_2>0$, provided the total heat capacity is negative, the resulting phase volume is (see Appendix \ref{app:C1+C2}):
\begin{align}
\Omega(E) \propto (-E)^{C_1+C_2} \quad C_1+C_2 <0
\label{eq:C1+C2}
\end{align}
This result resonates with the known fact that a positive and negative heat capacity systems can reach equilibrium when the negative heat capacity wins \cite{Thirring70ZPA235}.\footnote{Arguably equilibrium can be reached also under milder conditions. Ref. \cite{Campisi12EPL99} reported numerical evidence of equilibrium being reached between a single particle ($C_2=3/2$) in a box and a particle in the gravitational field ($C_1=-3/2$) for which $C_1+C_2=0$ } 

Now combine a macroscopic system with positive heat capacity with a macroscopic system with a slightly larger (in modulo) negative heat capacity. The result is a macroscopic system with a small (in modulo) negative heat capacity. 
Take for example a system of the type in Eq. (\ref{eq:HnegClarge}) with $f$ degrees of freedom, and an ideal gas with $f-k$ degrees of freedom, with $k$ some fixed integer, 
\begin{align}
H(\mathbf x,\mathbf p)&=\sum_{i=1}^{f} \frac{p_i^2}{2m} - \frac{G}{\sqrt{\sum_{i=1}^f{x_i^2}}} \nonumber \\
&+ \sum_{i=f+1}^{2f-k} \frac{p_i^2}{2m} + U_\text{box}^V(x_{f+1},\dots, x_{2f-k};V)
\label{eq:HnegC}
\end{align}
The resulting phase volume would be of the form
\begin{align}
\Omega(E) \propto (-E)^{-k/2}
\end{align}
independent of the value of $f$. For example for $k=3$, this would be the same phase volume as a single particle in a 3D gravitational field \cite{Thirring70ZPA235}, and  its shell-ergotropy, Eq. (\ref{eq:erg-}) would remain finite even in the limit of infinite $f$, thus avoiding the concentration of measure phenomeonon.
Despite the large number of degrees of freedom, the system in Eq. (\ref{eq:HnegC}), remains in a certain sense a \emph{small system}.

\section{General Case}
Equation (\ref{eq:erg+}) provides the shell-ergotropy of an ideal gas, or more generally of any system with constant positive heat capacity, like for example, a set of $N$ harmonic oscillators in $d$ dimensions for which $\gamma=dN=C$. 
The most important remark is that such shell-ergotropies exopnentially vanish in the large $\gamma$ limit for any fixed finite shell thickness $\epsilon$ and fixed energy $E$. In this regard it is worth mentioning the very recent results of \cite{Parrondo26PNASNEXUS8} presenting an estimate of the shell-ergotropy of the ideal gas, and reaching our same conclusion that it is vanishingly small. Similarly Eq. (\ref{eq:erg-}) gives the shell-ergotropy of systems with constant negative heat capacity.

As mentioned above, the vanishing of the shell-ergotropy in the {large $N$} limit is caused by the phenomenon of concentration of measure \cite{Levy51Book}. 
Clearly, this phenomenon may occur regardless of the shape of the hypersurfaces of constant energy, i.e., for a generic Hamiltonian $H$. 

Consider the case when the system admits a well defined \emph{thermodynamic limit}, in the sense that the 
phase volume obeys the scaling
\begin{align}
\Omega(E) \simeq \exp\left[N^\alpha s \left(\frac{E}{N^\beta}\right)\right],\quad \alpha,\beta>0
\label{eq:exponential-volume}
\end{align}
for some continuous function $s$ with non-null derivative, i.e., $s'>0$.\footnote{Note that Eq. (\ref{eq:exponential-volume}) holds for both systems with constant positive and negative heat capacities,  in which case it is $s(x)=\ln x, s(x)=-\ln(-x)$, respectively, and  $\alpha=\beta=1$.} Taking the logarithm, the thermodynamic entropy $\mathcal S= \ln \Omega$ \cite{Campisi25Book}, obeys
\begin{align}
\mathcal S(E) \simeq N^\alpha s \left({E}/{N^\beta}\right),\quad \alpha,\beta>0
\label{eq:entropy-scaling}
\end{align}
Note that from Eq. (\ref{eq:mu-S}) it is $\mathcal S(E')=\ln [\Omega(E)-\Omega(E-\epsilon)]$, thus we have, for  $\epsilon \ll E$
\begin{align}
\mathcal S(E') &= \mathcal S(E) + \ln \left [1 -\frac{\Omega(E-\epsilon)}{\Omega(E)} \right]\nonumber \\
& \simeq   \mathcal S(E) + \ln \left [1 -\exp\left( - N^{\alpha-\beta} s' ({E}/{N^\beta}) \epsilon \right)\right]
\label{eq:S(E')}
\end{align}
Since it is assumed that $s' > 0$, Eq. (\ref{eq:entropy-scaling}) can be inverted to give $E=N^\beta s^{-1}(\mathcal S/N^\alpha)$, therefore
\begin{align}
\frac{E'}{N^\beta} \simeq s^{-1}\left[s\left(\frac{E}{N^\beta}\right) + \frac{1}{N^\alpha} \ln \left (1 -\exp\left( - N^{\alpha-\beta} s' \left(\frac{E}{N^\beta}\right)  \epsilon \right)\right)
\right]
\label{eq:E'asymptotic1}
\end{align}
If we take the large $N$ limit while keeping $E\propto N^\beta$, we note that the second term in the square parentheses is negligible as compared to the first.\footnote{Looking at the second term in the square parentheses in Eq. (\ref{eq:E'asymptotic1}): i) when $\alpha=\beta$ the argument of the logarithm is of order $1$, so the term is of order $N^{-\alpha}$; ii) when $\alpha>\beta$ the argument of the logarithm approaches $1$, so the term vanishes faster than $N^{-\alpha}$; iii) when $\alpha < \beta$, the argument of the exponential is vanishing, thus expanding to first order, we see that the term  behaves as $N^{-\alpha}\ln (N^{\alpha-\beta} s' \varepsilon)$, i.e. it is still vanishing.} 
Expanding around $s\left({E}/{N^\beta}\right)$ we get, up to first order:
\begin{align}
\frac{E'}{N^\beta} \simeq \frac{E}{N^\beta} + \frac{1}{N^\alpha  s' }
\ln \left (1 -e^{ - N^{\alpha-\beta} s' \epsilon} \right)
\label{eq:E'asymptotic2}
\end{align} 
where we used $(s^{-1})'(x)= 1/(s'(s^{-1}(x))$, and used the light notation $s'$ for $s'(E/N^\beta)$. We thus get:
\begin{align}
\frac{E'-E}{E} \simeq \frac{1}{N^\alpha} \frac{1}{ s' E/N^\beta }\ln \left (1 -e^{ - N^{\alpha-\beta} s' \epsilon} \right)
\label{eq:E'asymptotic3}
\end{align}
This means that in ``thermodynamic limit'' ($N\to \infty$, $E/N^\beta=\text{const.}$) the relative
difference between $E$ and $E'$ is asymptotically vanishing. 
Therefore the rescaled ergotropy $\mathcal E/E$ is also asymptotically vanishing in the thermodynamic limit.
Note that it is crucial for this argument that $s' \neq 0$.

Let us also consider the mere ``large $N$'' limit (i.e. $N\to \infty$, $E=$const.). Taking the derivative of Eq. (\ref{eq:entropy-scaling}) with respect to energy, and recalling that it represents the inverse temperature, we get 
$T^{-1}\simeq N^{\alpha-\beta}s'$, namely 
\begin{align}
T(E,N)\simeq N^{\beta-\alpha}\vartheta(E/N^\beta),\quad \vartheta=1/s'
\label{eq:T}
\end{align}
With this Eq. (\ref{eq:S(E')}) reads $\mathcal S(E') = \mathcal S(E) +\ln(1-e^{-\varepsilon/T})$. Assuming that, at fixed $N$, $T$ vanishes with vanishing $E$ implies that $\vartheta(x) \to 0$, for $x\to 0$. Therefore for $\alpha \geq \beta$, $\mathcal S(E') - \mathcal S(E)$ vanishes for large $N$ and fixed $E$. Since $\mathcal S$ is continuous and invertible this means that $E-E'$ vanishes too. This continues to be so also for $\alpha<\beta$ if one further assumes that for small $x$, 
$
\vartheta(x)\propto x^\eta\ $ with $ \eta \geq 1
$,
which is a reasonable assumption. 

In sum, if $\alpha \geq \beta$ and $\vartheta(x)\to 0$ for small $x$, or if $\alpha < \beta$ and $\vartheta(x)\to 0$ at least linearly, not only it is expected that the relative ergotropy $\mathcal E/E$ vanishes in the ``thermodynamic limit'' (i.e. $N\to \infty$, $E/N^\beta=$const.), but also that the ergotropy itself $\mathcal E$ vanishes in the mere ``large $N$'' limit (i.e. $N\to \infty$, $E=$const.)

As an example, consider the ideal gas case. We have $\alpha = \beta =1$,  $s=\ln x, s'=1/x $, i.e., $s'>0$ on the domain $x>0$, while $\gamma$ plays the role of $N$. According to Eq. (\ref{eq:E'asymptotic3}) it is $(E'-E)/E\simeq (1/\gamma)\ln (1- e^{-\varepsilon \gamma/E})$, which is of order $1/\gamma$ in the thermodynamic limit ($\gamma \to \infty, E/\gamma$=const.). This is in agreement with the analytical expression in Eq. (\ref{eq:E'}). Note also that, in this case $T=E/\gamma$, so $\vartheta(x)=x$, which vanishes for small $x$, therefore $E-E'$ vanishes in the large $\gamma$ limit at fixed $E$, which is in agreement with the analytical calculations.


\section{Discussion}

Equation (\ref{eq:exponential-volume}) is known to be obeyed in a wide range of physically relevant cases including ordinary matter featuring short-range two-body interactions (for which $\alpha=\beta=1$), and as well systems with long-range interactions \cite{Campa09PR480}, 
(e.g., the long-range $XY$ model \cite{Toral04JSP114}, for which $\alpha=1,\beta=2$) and systems with non-local interactions (e.g.,  Thirring ``artificial model of a star'' \cite{Thirring70ZPA235,Campisi25Book}, for which $\alpha=1,\beta=2$, but the conditions of contuinity and monotonicity of $s$ are not met everywhere).

It is noteworthy that, whenever Eq. (\ref{eq:exponential-volume}) holds, the thermodynamic entropy can be equivalently expressed as the logarithm of the phase volume $\Omega$ (as prescribed by Helmholtz theorem \cite{Campisi05SHPMP36,Campisi25Book}), or the logarithm of the density of states $\omega=\Omega'$, because the difference in their predictions of thermodynamic quantities (such as temperature, pressure, magnetisation etc.) would be negligible \cite{Campisi25Book}. We thus see that the problem of finding classical systems whose shell-ergotropy does not vanish in the {large $N$} limit is strictly linked  to that of  finding classical systems for which the so called ``volume entropy'' ($\ln \Omega$)  and ``surface entropy'' ($\ln \omega$) are not equivalent. Whether one or the other is the fundamentally correct expression of thermodynamic entropy has been highly debated in the literature \cite{Dunkel14NATPHYS10,Hilbert14PRE90,Campisi15PRE91}, while so far, examples of large \emph{classical} Hamiltonian system  where such equivalence breaks have remained elusive. Our example above, Eq.  (\ref{eq:HnegC}) featuring a large system with a small negative heat capacity, fills that gap, providing at the same time an example of a large classical system with non-vanishing shell-ergotropy.
It must be remarked that such model is rather artificial. In particular note the rather un-physical $N$-body interaction term in Eq. (\ref{eq:HnegC}).

Furthermore, we remark that any state of the form $\rho=F(H)$, with $F(x)$ a generic limited function (possibly featuring jumps and flat plateaus) can be approximated to any wanted degree,\footnote{E.g., in the sense of vanishing $L^1$ norm distance.} by a convex linear combination of shell-distributions. Thus, under the provision of the scaling in Eq. (\ref{eq:exponential-volume}), and the further assumption noted below Eq. (\ref{eq:T}), any cyclical driving would map (in the {large $N$} limit) each shell to a set of larger or same energy expectation,  so that the full distribution $\rho$ would get mapped to a new distribution with higher or same energy. It follows that it is
\begin{align}
\lim_{N \to \infty}\mathcal E[F(H)] = 0
\label{eq:erg-fate}
\end{align}
namely, under the provision of the stated assumptions, any stationary state of the form $F(H)$ is asymptotically passive. Note the crucial fact that $F$ does not need to be non-increasing. We remark that the limit in Eq. (\ref{eq:erg-fate}) must not be understood as being taken while rescaling the system average energy.

An interesting consequence of Eq. (\ref{eq:erg-fate}) is that ergotropy tends to become fully coherent in the {large $N$} limit. As was proved in \cite{Campisi26EPL154}, ergotropy can be split into a coherent and an incoherent part $\mathcal E=\mathcal E_c+\mathcal E_i$ (a fact that was first evidenced in the quantum realm \cite{Francica20PRL125}) where the incoherent part $\mathcal E_i[\rho]$ is defined as the ergotropy of the decohered density $\mathcal D[\rho]$, i.e., $\mathcal E_i[\rho]=\mathcal E [\mathcal D[\rho]]$. In the classical context the decoherence operator $\mathcal D$ is the operator that homogenises the distribution $\rho$ over the hyper-surfaces of constant energy $H$ \cite{Smith22entropy22,Campisi26EPL154}. As a consequence the decohered density $\mathcal D[\rho]$ is of the form $\mathcal D[\rho]=F(H)$ \cite{Campisi26EPL154}. Then, by virtue of Eq. (\ref{eq:erg-fate}), the incoherent ergotropy is asymptotically vanishing, 
$
\lim_{N \to \infty}\mathcal E_i = 0
$
that is, under the provision of Eq. (\ref{eq:exponential-volume}) and the further assumption noted below Eq. (\ref{eq:T}), the ergotropy is fully coherent in the {large $N$} limit
\begin{align}
\mathcal E \simeq \mathcal E_c, \quad \text{for } N\gg1
\end{align}
This is expected to occur as well in the quantum case.

It is noteworthy that our finding of Eq. (\ref{eq:erg-fate}) might look, at first sight, to be in contrast with the known fact that when considering many non-interacting copies of a system, increasing their number activates the total compound system, unless the system is in a thermal state (which is completely passive). This is known for both classical \cite{Gorecki80LMP4} and quantum systems \cite{Alicki13PRE87}. Thus, in the large $N$ limit, an $N$-copy compound system typically tends to become active. This is not in disagreement with our finding because, while the compound state of the $N$-copy system 
$ 
\rho_\text{tot}(\mathbf z) =\rho(\mathbf z_1) \rho(\mathbf z_2) \dots \rho(\mathbf z_N)
$ 
 is a stationary state relative to the total Hamiltonian 
$ 
H_\text{tot}(\mathbf z)= H(\mathbf z_1) + H(\mathbf z_2)+ \dots +H(\mathbf z_N) 
$ 
it is generally not of the form $F( H_\text{tot}(\mathbf z))$.

Finally, let us discuss the relation between the results presented here and those presented of Ref. \cite{Allahverdyan11EPL95}. That work considers the ergotropy of $N$ spins $1/2$ with energy gap $\delta$, being prepared in a flat mixture of all states belonging to a given degenerate eigenenergy. 
In that case the ergotropy is found to be of the order of the gap $\delta$, for eigenenergies in the lower half of the energy spectrum.  The discrepancy with the behaviour of the shell-ergotropy evidenced above originates from the discreteness of the spectrum, as compared to the classical contimuous case. Because of said discreteness, regardless of how large the spin system system is, there is always at least some spin in the considered state that can be flipped to decrease the energy, hence its ergotropy cannot tend to zero unless the system is prepared in the ground state.  Note however that it does so in the limit of vanishing gap $\delta$ where the spectrum approaches the continuum limit. 
Ref. \cite{Allahverdyan11EPL95} also reported that the ergotropy scales sub-extensively, specifically as $\delta \sqrt{N\ln N}$, for the single eigenenergy exactly at the middle of the energy spectrum. Such state corresponds to a point where the derivative of the phase volume vanishes \cite{Campisi15PRE91}, i.e., where the function $s' \propto \omega/\Omega$ (i.e., the inverse temperature) vanishes, and, accordingly,  the r.h.s. of Eq.  (\ref{eq:E'asymptotic3}) diverges. This results in a breakdown of the equivalence of ``surface'' and ``volume'' entropies (as evidenced previously in Ref. \cite{Campisi15PRE91}) and the according non-vaninshing of the shell-ergotropy. 
We note that such states of infinite (microcanonical) temperature can only exists for systems with bounded spectrum, which excludes systems with a kinetic term in the Hamiltonian. This clarifies why the results of Ref. \cite{Allahverdyan11EPL95}  were not in full agreement with Refs. \cite{Tasaki00arXivb,Allahverdyan02PHYSA305}.

\section{Conclusions}
We have established a connection between the physical problem of finding the available energy, a.k.a. ergotropy, of a classical system and the mathematical theory of function rearrangement in $\mathbb R^N$. By noting that the ``ergotropic rearrangement'' is a generalisation of the classic concept of ``symmetric decreasing rearrangement'', one easily finds an expression of the minimal energy state $\breve \rho$, valid under most general assumptions, including the case when the system state $\rho$ presents jump discontinuities and flat regions, thus extending the results previously obtained in Ref. \cite{Campisi26EPL154}.

In ``ergotropic rearrangement'' a phase space density $\rho_0(\mathbf z)$ is rearranged in decreasing manner relative to  the increase of the Hamilton function $H(\mathbf{z})$. Clearly the method does not require that $H$ is a Hamilton function, nor that the space $\mathbf z$ is a phase-space. In other words we have illustrated a method to rearrange any positive measurable function $p:\mathbb R^N \to \mathbb R_+$ in a decreasing manner relative to a generic function $h:\mathbb R^N \to \mathbb R$. Such "generalised rearrangements" can well find applications beyond the study of ergotropy. 

As an application, we have used the ``ergotropic rearrangement'' formula to study an important fundamental question, namely what is  the fate of ergotropy in the {large $N$} limit. Our main conclusion, Eq. (\ref{eq:erg-fate}), is that any equilibrium state of the form $F(H)$, is asymptotically passive, provided the ubiquitous scaling (\ref{eq:exponential-volume}) applies together with the further assumption noted below Eq. (\ref{eq:T}). We have illustrated this with macroscopic systems with constant (either positive or negative) heat capacity. 

We further noticed that said scaling condition also implies the equivalence of ``surface'' and ``volume'' entropies'' \cite{Dunkel14NATPHYS10,Hilbert14PRE90,Campisi15PRE91} thus suggesting that ergotropy extraction from a microcanonical ensemble should be looked for in systems where such equivalence breaks. This is known to be the case for small systems, but no example where such breakdown occurs had been reported so far in the case of a large classical system. Here we provided one such (somewhat artificial) example featuring both inequivalence of surface and volume entropies, and non vanishing shell-ergotorpy.

Summing up, if you have a classical macroscopic ergodic system obeying the scaling (\ref{eq:exponential-volume}) of which you only know it is exploring some energy hypersurface, even knowing its energy with any finite accuracy, will not help you extract any energy, except for possibly very peculiar cases that do not obey the further assumptions discussed above. On the contrary, you should be able to do so in low dimensional systems, as was in fact demonstrated in various works  \cite{Sato02JPSJ71,Marathe10PRL104,Vaikuntanathan11PRE83,Soriani19JSM19}, and the energy extraction becomes harder and harder as the dimensionality increases.

These findings further elucidate the mechanical foundations of Kelvin formulation of the second Law of Thermodynamics \cite{Allahverdyan02PHYSA305,Campisi25Book} by evidencing that \emph{smallness} (either due to a small number of degrees of freedom, or to the absence of the scaling condition  (\ref{eq:exponential-volume}) in case of  many degrees of freedom) is necessary for microcanonical Szilard engines to work \cite{Sato02JPSJ71,Marathe10PRL104,Vaikuntanathan11PRE83,Soriani19JSM19}.

\subsection*{Use of artificial intelligence}
All the material and results presented were solely produced by the author, without any aid of artificial intelligence.

\appendix

\section{Asymptotic expansions}
\label{app:asymptotics}
To find the asymptotic (large $\gamma$) expansion of $f_\gamma(x)$, Eq. (\ref{eq:f-gamma}), let us first expand the function
\begin{align}
a_\gamma(y)\doteq \frac{1-y^{\gamma+1}}{1-y^\gamma}, \quad y\in(0,1)
\end{align}
where, for simplicity we introduced the quantity $y=1-x$. 
Using the expansion $1/(1-s)\simeq 1+s$ for $s\ll1$, and noting that $y^\gamma \ll1$ for large $\gamma$, we have 
\begin{align}
a_\gamma(y) \simeq (1-y^{\gamma+1}) (1+y^\gamma)
\simeq1+y^\gamma(1-y)
\end{align}
where we noted that term $y^{2\gamma+1}$ is negligible as compared to the others.

Let us now expand the function:
\begin{align}
b_\gamma(y)\doteq [1-y^\gamma]^{1/\gamma} , \quad y\in(0,1)
\end{align}
Using the small $s$ expansion, $(1-s)^a \simeq 1-a s$, we get
\begin{align}
b_\gamma(y)\simeq 1- y^\gamma/\gamma
\end{align}
We thus get
\begin{align}
f_\gamma(y)= \frac{\gamma}{\gamma+1} [a_\gamma(y)-b_\gamma(x)] \simeq y^\gamma(1-y)
\end{align}
where we used that $\gamma/(\gamma+1)\simeq 1$ and that $y^\gamma/\gamma$ is negligible as compared to $y^\gamma(1-y)$. Recalling that $y=1-x$, leads to Eq. (\ref{eq:f-asymptotic}).

Let us now turn to the asymptotic expansion of $g_\gamma(x)$, Eq. (\ref{eq:g-gamma}). Consider the term
\begin{align}
c_\gamma(y)\doteq \frac{y^{1-\gamma}-1}{y^{-\gamma}-1}, \quad y\in(0,1)
\end{align}
Multiplying and dividing by $y^\gamma$, we get:
\begin{align}
c_\gamma(y)&= y \frac{1-y^{\gamma-1}}{1-y^{\gamma}}
\simeq y (1-y^{\gamma-1})(1+y^{\gamma}) \nonumber \\
&\simeq y [1-y^{\gamma-1}(1-y)] = y -y^\gamma(1-y)
\end{align}
Turning to the term:
\begin{align}
d_\gamma(y)\doteq [y^{-\gamma}-1]^{-1/\gamma} , \quad y\in(0,1)
\end{align}
we have,
\begin{align}
d_\gamma(y)= y [1-y^{\gamma}]^{-1/\gamma} \simeq y+y^{\gamma+1}/\gamma 
\end{align}
Summing up
\begin{align}
g_\gamma(y)=  \frac{\gamma}{1-\gamma}  [c_\gamma(y)-d_\gamma(x)] \simeq y^\gamma(1-y)
\end{align}
where, just like in the previous case, the term $y^{\gamma+1}/\gamma$ can be neglected, and we used  $\gamma/(1-\gamma) \simeq -1$.
Eq. (\ref{eq:g-asymptotic}) follows from the substitution $y=1-x$.

\section{Phase volume of the Hamiltonian in Eq. (\ref{eq:HnegClarge})}
\label{app:phase-volume-negC}
Using the definition of phase volume, Eq. (\ref{eq:Omega_0}), and using the short notation $\mathbf p = (p_1 \dots p_f), \mathbf x = (x_1 \dots x_f)$ we have:
\begin{align}
\Omega(E)= \int d\mathbf x d\mathbf p\,  \theta(E+G/|\mathbf x|- \mathbf p^2/(2m))
\end{align}
Performing the integration in $d \mathbf p$ and noting that it amounts to the measure of a hypersphere of radius $R=\sqrt{2m(E-G/|\mathbf x|)}$ in $f$ dimensions we get
\begin{align}
\Omega(E) \propto \int d\mathbf x [E + G/|\mathbf x|]_+^{f/2}
\end{align}
where $[x]_+=x\theta(x)$. Turning now to hyperspherical coordinates, we get
\begin{align}
\Omega(E) &\propto \int_0^{-G/E} dr\, r^{f-1} [E + G/r]^{f/2} \nonumber \\
&= (-E)^{f/2}  \int_0^{-G/E} dr\, r^{f-1} [-1 + G/(-rE)]^{f/2}  \nonumber
\end{align}
where we recall that in order for the system to remain confined the condition $E<0$ is necessary, otherwise the phase volume would diverge.
Making the change of variable $y=-rE/G$, we get
\begin{align}
\Omega(E) &\propto  (-E)^{-f/2} \int_0^{1} dy\, y^{f-1} [1/y-1]^{f/2} 
\end{align}
with the integral being well defined for any positive integer $f$.

\bigskip
\section{Derivation of Eq. (\ref{eq:erg-})}
\label{app:derivation-erg-}
We recall that in the case of negative heat capacity systems, the energy range is $E\in(-\infty, 0)$. For positive energies no bound states exist, while the energy tends to $-\infty$ as one approaches the origin of phase space, see Eq. (\ref{eq:HnegClarge}).

For convenience, the boundaries of the shell $S$ are now set to $E$ and $E+\epsilon$.
The measure of the shell $S=\{\mathbf x | E<H(\mathbf x) < E+\epsilon \}$ is:
\begin{align}
\mu(S)= B_\gamma [(-E-\epsilon)^{-\gamma}-(-E)^{1/\gamma}]
\end{align}
Inverting Eq. (\ref{eq:Omega-negC}) we have:
\begin{align}
\Omega^{-1}(x) = -\left(\frac{x}{B_\gamma}\right)^{-1/\gamma}
\end{align}
The bound energy $E'$, Eq. (\ref{eq:E'-def}), of the ergotropic rearrangement, $\breve S$,  is then:
\begin{align}
E'= - [(-E-\epsilon)^{-\gamma}-(-E)^{1/\gamma}]^{-1/\gamma}
\end{align}
For the initial energy expectation we have:
\begin{align}
\bar E&
= \frac{1}{\mu(S)} \int_{E}^{E+\epsilon} de\,  \omega(e) e 
= \frac{\gamma B_\gamma}{\mu(S)} \int_{E}^{E+\epsilon} de\,  (-e)^{-\gamma-1} e \nonumber  \\
&= \frac{\gamma B_\gamma}{\mu(S)} \int^{-E-\epsilon}_{-E} dy\,  y^{-\gamma}  \nonumber  \\
&=\frac{B_\gamma}{\mu(S)} \frac{\gamma}{1-\gamma} \left[ y^{1-\gamma} \right]^{-E-\epsilon}_{-E}  \quad  \quad (\text{for } \gamma \neq1)
 \nonumber  \\
&= \frac{\gamma}{1-\gamma}\frac{(-E-\epsilon)^{1-\gamma}-(-E)^{1-\gamma}}{(-E-\epsilon)^{-\gamma}-(-E)^{-\gamma}} 
\end{align}
where $\omega(e)=d\Omega(e)/de= \gamma B_\gamma (-e)^{-\gamma-1}$ is the density of states.
Similarly, the lowest energy expectation the system can reach is:
\begin{align}
\breve E
&= \frac{\gamma B_\gamma}{\mu(S)} \int^{E'}_{-\infty} de\,  (-e)^{-\gamma-1}e
=\frac{\gamma B_\gamma}{\mu(S)} \int^{-E'}_{+\infty} dy\,  y^{-\gamma}\nonumber \\
&=\frac{B_\gamma}{\mu(S)} \frac{\gamma}{1-\gamma} \left[ y^{1-\gamma} \right]^{-E'}_{+\infty} \quad  \quad (\text{for } \gamma>1)
 \nonumber  \\
 &=  \frac{\gamma}{1-\gamma} 
 [(-E-\epsilon)^{-\gamma}-(-E)^{-\gamma}]^{-1/\gamma}
\end{align}
where we used that $E_0=-\infty$.

After some simple manipulations one arrives at Eq. (\ref{eq:erg-}).

Note that the case $\gamma=1$ should be treated separately. We leave this as en exercise for the interested reader.

\section{Combining a constant positive heat capacity system with one with constant negative heat capacity}
\label{app:C1+C2}
To prove Eq. (\ref{eq:C1+C2}), consider the total Hamiltonian
\begin{align}
H(\mathbf z_1,\mathbf z_2)= H_1(\mathbf z_1) +H_2(\mathbf z_2)
\end{align}
where $H_{1(2)}$ is the Hamiltonian of the system with constant negative(positive) heat capacity.
Let's apply Eq. (\ref{eq:Omega_0}) to calculate the total system phase volume:
\begin{align}
\Omega(E) = \int d\mathbf z_1 d\mathbf z_2 \theta(E-H_1(\mathbf z_1) -H_2(\mathbf z_2))
\end{align}
Performing the integration over $d\mathbf z_2$ we get
\begin{align}
\Omega(E) \propto \int d\mathbf z_1 [E-H_1(\mathbf z_1)]_+^{C_2}
\end{align}
where $[x]_+\doteq x\theta(x) $ and we have applied Eq. (\ref{eq:Omega_0}) to system $2$ using that its phase volume $\Omega_2$ is proportional to $E^{C_2}$ (and is defined for $E>0$). Noting that the integrand depends on $\mathbf z_1$ via the Hamiltonian $H_1$ allows to write the phase space integral as an energy integral using  of system-1's density of states $\omega_1=\Omega_1'\propto (-E)^{C_1-1}$ \cite{Campisi25Book}, i.e.,
\begin{align}
\Omega(E) \propto \int_{-\infty}^E dE_1 (-E_1)^{C_1} (E-E_1)^{C_2-1}
\end{align}
The dependence on $E$ can be easily isolated via the change of variable $y=E_1/E$ (with $E<0$)
\begin{align}
\Omega(E) \propto (-E)^{C_2+C_1}\int_{1}^\infty dy \,y^{C_1} (y-1)^{C_2-1}
\end{align}
and the integral converges provided $C_1+C_2<0$, which concludes the proof.

%

\end{document}